\documentclass[aps,pre,twocolumn,showpacs,epsfig,floats]{revtex4}
\usepackage{amsmath}
\usepackage{color}
\usepackage{graphicx}
\usepackage{epsfig}
\usepackage{amsfonts}
\usepackage{mathrsfs}
\usepackage{mathtools}
\usepackage{amsmath}
\usepackage{float}
\usepackage{physics}

\begin{document} 

\title{Chaos and quantum scars in coupled top model}
 
\author{Debabrata Mondal, Sudip Sinha and S. Sinha}
\affiliation{Indian Institute of Science Education and
Research-Kolkata, Mohanpur, Nadia-741246, India}
 
\date{\today}

\begin{abstract}
We consider a coupled top model describing two interacting large spins, which is studied semiclassically as well as quantum mechanically. This model exhibits variety of interesting phenomena such as quantum phase transition (QPT), dynamical transition and excited state quantum phase transitions above a critical coupling strength. Both classical dynamics and entanglement entropy reveals ergodic behavior at the center of energy density band for an intermediate range of coupling strength above QPT, where the level spacing distribution changes from Poissonian to Wigner-Dyson statistics. Interestingly, in this model we identify quantum scars as reminiscence of unstable collective dynamics even in presence of interaction. Statistical properties of such scarred states deviate from ergodic limit corresponding to random matrix theory and violate Berry's conjecture. In contrast to ergodic evolution, oscillatory behavior in dynamics of unequal time commutator and survival probability is observed as dynamical signature of quantum scar, which can be relevant for its detection. 
\end{abstract}


\maketitle

{\it Introduction:}
After the recent experiment on a chain of Rydberg atoms \cite{Rydberg}, quantum scar (QS) in many body systems has drawn significant interest due to its connection with ergodicity and non equilibrium dynamics. Observed athermal behavior and periodic revival phenomena in dynamics of a specific initial state has been attributed to the  many body quantum scarring phenomena \cite{Turner,Turner2_PRB,Turner3_PRX,Lin,rydberg_2d,floquet1_krishnendu,floquet2_kaoru}. Many body quantum scar (MBQS) gives rise to the deviation from ergodicity \cite{Turner,Turner2_PRB,Turner3_PRX,Lin,rydberg_2d,floquet1_krishnendu,floquet2_kaoru,
ETH_violation_AKLT,ETH_violation_driven,Onsager_scars,Sanjay_MPS,spin_1_xy,
emergent_SU2_Choi,choi_2,Sanjay_2,AKLT2_D_Mark,AKLT_chains,Changlani,ising_ladder,eta_pairing,driven_fracton} leading to the violation of eigenstate thermalization hypothesis (ETH) \cite{ETH} in quantum systems \cite{floquet2_kaoru,ETH_violation_AKLT,Lin,ETH_violation_driven,Onsager_scars,Sanjay_MPS}. Recent theoretical studies on interacting spin systems reveal that emergent symmetries and symmetry protected many body states are the key ingredient for MBQS \cite{Onsager_scars,spin_1_xy,emergent_SU2_Choi,choi_2,Sanjay_2,AKLT2_D_Mark,AKLT_chains,
Changlani,ising_ladder,eta_pairing}. Originally QS in non interacting quantum chaotic system has been identified as reminiscence of {\it unstable} classical orbits \cite{Heller,Kaplan}. 
However such connection between MBQS and underlying dynamics remains unclear for interacting many body systems. Similar problem also arises in understanding ergodicity in quantum many particle systems from the viewpoint of chaos and phase space mixing. However some studies on quantum collective models reveal a connection between ergodicity and underlying chaotic dynamics \cite{dicke1,dicke2,dicke3,Mori}, as well elucidate the role of underlying unstable collective dynamics in the formation of MBQS \cite{sinha}.

In this work we consider coupled top model which allows us to study the collective dynamics of two large spins semiclassically, revealing interesting phenomena like quantum phase transition (QPT), dynamical transition (DT) \cite{DT_expt1,Oberthaler,shenoy,note0}, and chaotic behavior at an intermediate regime of coupling strength. More importantly, in the chaotic regime, we identify QS as a signature of unstable orbits of collective spin dynamics within certain symmetry classes. Statistical and dynamical properties of such scarred states exhibit clear deviation from ergodicity.
  
{\it Model and semiclassical analysis:} The coupled top model \cite{chaos,kicked_spin1,classical_analysis,ct_EE,kicked_spin2} describes the dynamics of ferromagnetically coupled two large spins similar to the transverse field Ising model \cite{q_ising1,q_ising2}, which is governed by the Hamiltonian,
\begin{equation}
\hat{\mathcal{H}} =-\hat{S}_{1x} -\hat{S}_{2x}-\frac{\mu}{S}\hat{S}_{1z}\hat{S}_{2z},
\label{coupled_top_model}
\end{equation}
where $\hat{S}_{ia}$ represents components $(a=x,y,z)$ of two large spins of equal magnitude $S$ denoted by index $i=1,2$ and $\mu$ is the ferromagnetic coupling strength. 

\begin{figure}[H]
    \centering
    \includegraphics[height=6.1cm,width=9.05cm]{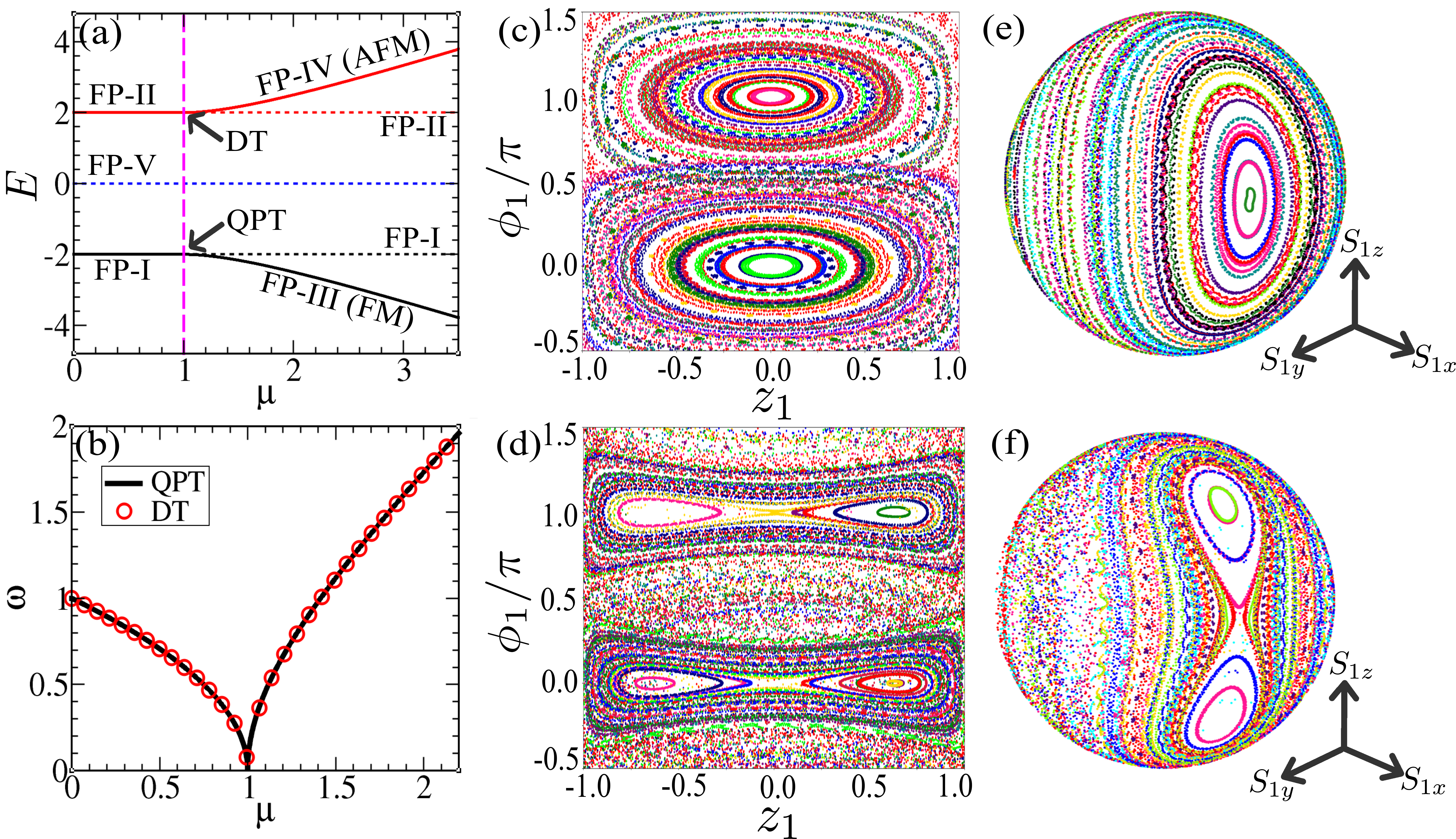}
    \caption{(a) Schematic diagram of different steady states at different energies with increasing $\mu$. Solid (dotted) lines represent stable (unstable) branches and vertical pink dashed line marks critical coupling $\mu_c$. (b) Small oscillation frequency as a function of $\mu$ for QPT (black solid line) and DT (red circles). Phase portraits ((c) and (d)) and trajectories on Bloch sphere ((e) and (f)) exhibiting QPT and DT. For (c),(e) $\mu=0.8< \mu_c$ and (d),(f) $\mu=1.3>\mu_c$.}
    \label{fig:1}
\end{figure}

Large magnitude of spin $S\gg 1$ allows us to analyze the model semiclassically, where the spin vectors are represented by $\vec{S}_{i}=(S\sin\theta_i\cos\phi_i,S\sin\theta_i\sin\phi_i,S\cos\theta_i)$. In terms of these dynamical variables, corresponding classical Hamiltonian can be written as,
\begin{equation}
\mathcal{H}_{cl}=-\sqrt{1-{z_1}^2}\cos\phi_1-\sqrt{1-{z_2}^2}\cos\phi_2-\mu \,z_1 z_2
\label{class_Ham}
\end{equation}
where $z_i = \cos\theta_i$ is conjugate momentum corresponding to the variable $\phi_i$. Here $\mathcal{H}_{cl}$ and classical energy $E$ are scaled by $S$. The classical spin dynamics is described by following equations of motion (EOM),
\begin{eqnarray}
\dot{z_i}=-\sqrt{1-{z_i}^2}\sin\phi_i\,; \quad \dot{\phi_i}=\frac{z_i\cos\phi_i}{\sqrt{1-{z_i}^2}}-\mu z_{\bar{i}} \label{EOM}
\end{eqnarray} 
where $\bar{i} \neq i$. To understand the overall behavior of dynamics, we first analyze the fixed points (FP) and their stability for varying coupling strength $\mu$, as charted in Fig.\ref{fig:1}(a). Due to the interaction between the spins, at a critical coupling strength $\mu_c =1$, the model undergoes a QPT to ferromagnetically ordered ground state (GS) as well as dynamical transition corresponding to the highest excited state (ES) with anti-ferromagnetic ordering. For $\mu < \mu_c$, two symmetry unbroken stable FPs are represented by: (I) $\{z_1=0,\phi_1=0,z_2=0,\phi_2=0\}$ with energy density $E=-2$ (GS) and (II) $\{z_1=0,\phi_1=\pi,z_2=0,\phi_2=\pi\}$ with $E=2$ (ES). 
For symmetry unbroken phase FP-I (FP-II), both spins are aligned to positive (negative) x-axis without any magnetization along z-axis. Both FP-I and II undergoes a pitchfork bifurcation at $\mu_c$ and become unstable for $\mu>\mu_c$. As a result of bifurcation, two pairs of symmetry broken stable steady states appear above $\mu_c$, namely: (III) 
$\{z_1=z_2=\pm\sqrt{1-1/\mu^2},\phi_1= \phi_2=0\}$ with $E=-(\mu+1/\mu)$ which corresponds to the ferromagnetic GS and (IV) $\{z_1=-z_2=\pm\sqrt{1-1/\mu^2},\phi_1=\phi_2=\pi\}$ representing a dynamically stable anti-ferromagnetic state with $E=(\mu+1/\mu)$ corresponding to the ES. Apart from these states, there exists another pair of unstable FPs, which we denote as `$\pi$-mode', describing the relative angle between the spins: (V) $\{z_1=z_2=0,\phi_1=\pi,\phi_2=0\}$ and $\{z_1=z_2=0,\phi_1=0,\phi_2=\pi\}$ with $E=0$.

Interestingly, the dynamical transition corresponding to ES gives rise to the anti-ferromagnetic spin configuration, which survives as a dynamically stable steady state (FP-IV) in spite of ferromagnetic interaction between them. This can be understood from the transformation $\hat{S}_{x}\rightarrow -\hat{S}_{x}$ and $\mu \rightarrow -\mu$ in Eq.(\ref{coupled_top_model}), which converts the ES to GS of anti-ferromagnetic coupled top model. As shown in Fig.\ref{fig:1}(b), small amplitude oscillation frequencies (excitation frequencies) corresponding to both GS and ES vanishes as $\sim \sqrt{|\mu - \mu_c|}$ at the critical point, signifying the QPT and dynamical transition. Both the transitions can also be confirmed from full quantum analysis of the model \cite{ct_EE,sm}.

To gain more information, we study phase space trajectories over the Bloch sphere and phase portrait in the $z-\phi$ plane by solving Eq.(\ref{EOM}), as depicted in Fig.\ref{fig:1}(c)-(f).
The bifurcation of FP (from I to III) corresponding to the QPT is evident from the phase space trajectories on the Bloch sphere (see Fig.\ref{fig:1}(e),(f)). 
For $\mu<\mu_c$, the phase portrait in the $z-\phi$ plane contains regular trajectories around the symmetry unbroken FPs with $z=0$, as shown in Fig.\ref{fig:1}(c). Whereas above the critical point, the symmetry broken FPs with $z\neq 0$ in the phase portrait correspond to QPT and dynamical transition, however intermediate region of phase space is filled with irregular (chaotic) trajectories (see Fig.\ref{fig:1}(d)). Such mixed phase space structure indicates interesting ergodic behavior \cite{Turner3_PRX,lewenstein,zaslavasky_book}, which leads us to further study the model quantum mechanically.

{\it Quantum chaos and ergodicity:}
To analyze the system quantum mechanically, we diagonalize the Hamiltonian Eq.(\ref{coupled_top_model}) for a fixed magnitude of spin $S$ so that the dimensionality of Hilbert space becomes $\mathcal{N} = (2S +1)^2$.
The eigenvalues $\mathcal{E}_n$ and corresponding eigenfunctions $\ket{\psi_n}$ of the Hamiltonian are obtained using the basis states $\ket{m_{1z},m_{2z}}$, where $m_{iz}$ is eigenvalue of $\hat{S}_{iz}$.
The eigenfunctions and eigenvalues can be classified according to the symmetry of the Hamiltonian (Eq.(\ref{coupled_top_model})) which remains invariant under parity $\hat{\Pi}=e^{-\imath\pi\hat{\mathcal{P}}}$ where $\hat{\mathcal{P}} = \hat{S}_{1x} +\hat{S}_{2x}$ and under exchange symmetry between two spins ($S_{1}\leftrightarrow S_{2}$) \cite{note1}. From the quantum mechanical analysis of the model, we notice for $\mu>\mu_c$, both QPT and dynamical transition lead to excited state quantum phase transitions (ESQPT) \cite{Cejnar1,Caprio,Cejnar2,dicke_ESQPT,L_Santos} corresponding to the unstable symmetry unbroken FPs I and II with energy densities $E=-2$ and $2$, respectively, where the derivative of semiclassical density of state becomes singular \cite{sm,Cejnar1,Caprio,Cejnar2,dicke_ESQPT,L_Santos}.

\begin{figure} [H]
    \centering
    \includegraphics[height=7.7cm,width=8.7cm]{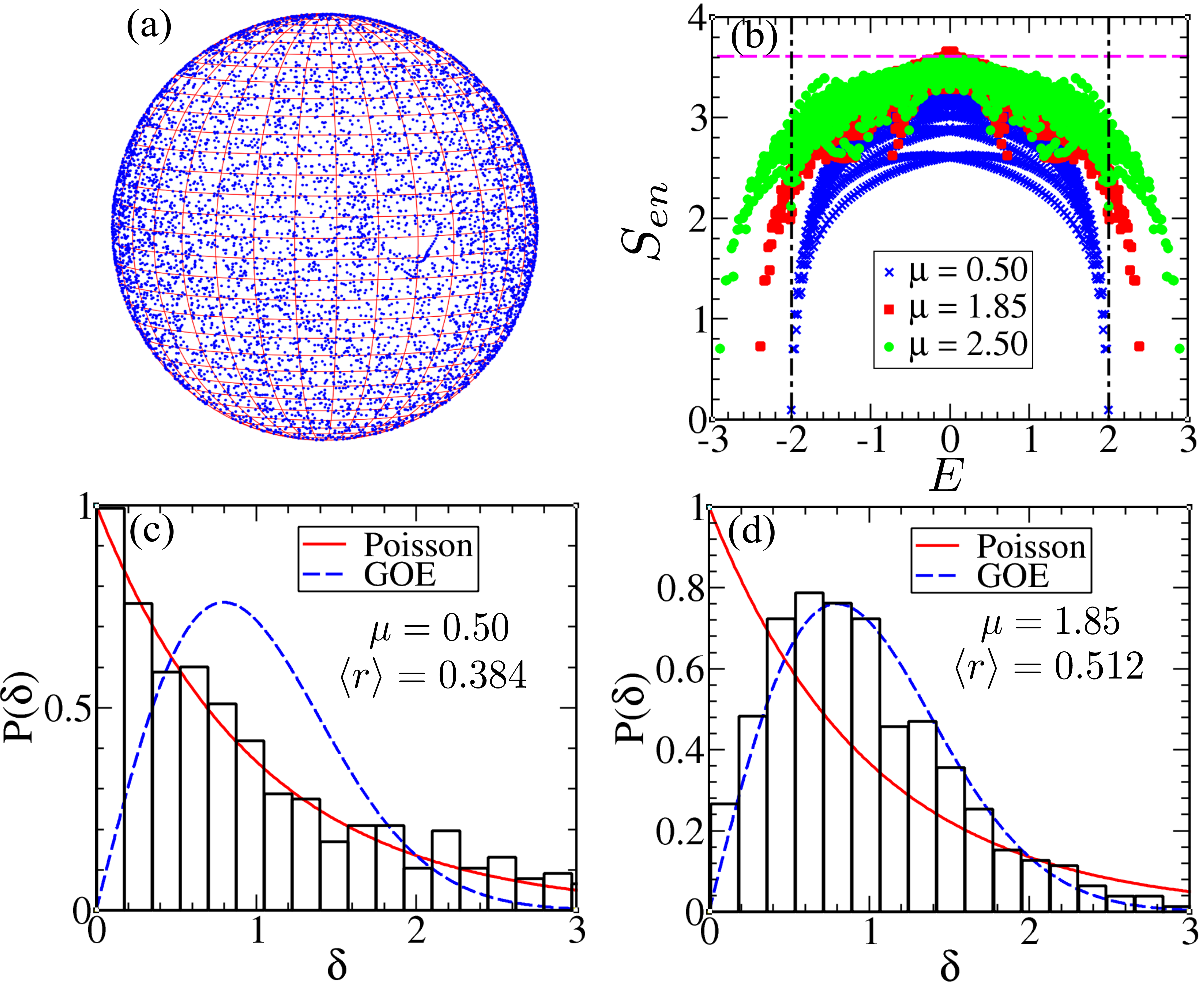}
    \caption{(a) Chaotic behaviour of a trajectory on Bloch sphere corresponding to $E\approx 0$ and $\mu=1.85$. (b) Variation of EE of eigenstates with energy density $E$ for different $\mu$. Horizontal pink dashed line represents the maximum EE corresponding to Eq.(\ref{max_EE}) and vertical black dashed-dotted lines indicate ESQPT. Level spacing distribution for (c) regular and (d) ergodic regime.
Parameters chosen: $S = 30$ for (b), (c), (d) and all other figures unless otherwise mentioned.}
    \label{fig:2}
\end{figure}

Next, we study the entanglement entropy (EE) of the eigenstates which contains useful information related to ergodicity, as it is expected that chaoticity and phase space mixing can lead to enhancement of EE \cite{sm,lewenstein,shohini_ghose,expt_lin}. The reduced density matrix of {\scriptsize $\mathcal{S}$}-th spin  $\hat{\rho}_{\mathcal{S}} =\text{Tr}_{\bar{\mathcal{S}}}\ket{\psi_n}\bra{\psi_n}$ is obtained by tracing out other spin sector ({\scriptsize $\bar{\mathcal{S}} \neq \mathcal{S}$}) which yields the EE, $S_{en} = -\text{Tr}(\hat{\rho}_{\mathcal{S}}\text{log}\hat{\rho}_\mathcal{S})$ of the corresponding state $\ket{\psi_n}$. We quantify the degree of ergodicity of a state by comparing it with the EE of maximally random state partitioned into subsystems A(B) with dimension $\mathcal{D}_A$($\mathcal{D}_B$). Maximal EE corresponding to subsystem A with $D_A\leq D_B$ \cite{Page} is given by, 
\begin{equation}
S_{\text{max}}\simeq \text{log}(\mathcal{D}_A)-\mathcal{D}_A/2\mathcal{D}_B 
\label{max_EE}
\end{equation} 
where $\mathcal{D}_A=\mathcal{D}_B=2S+1$. 
For any value of coupling strength $\mu$, the eigenstates with maximum EE are located at the band center with energy density $E\approx 0$, as shown in Fig.\ref{fig:2}(b). Moreover, EE corresponding to the band center ($E\approx 0$), attains its maximum possible limit $S_{\text{max}}$ (given in Eq.(\ref{max_EE})) around $\mu \approx 2$, indicating ergodic behavior. In this region, we also observe classically chaotic phase space trajectories with classical energy $E\approx 0$, which fill up the Bloch sphere completely, as depicted in Fig.\ref{fig:2}(a).
 
\begin{figure}[H]
    \centering
    \includegraphics[height=7.7cm,width=9.1cm]{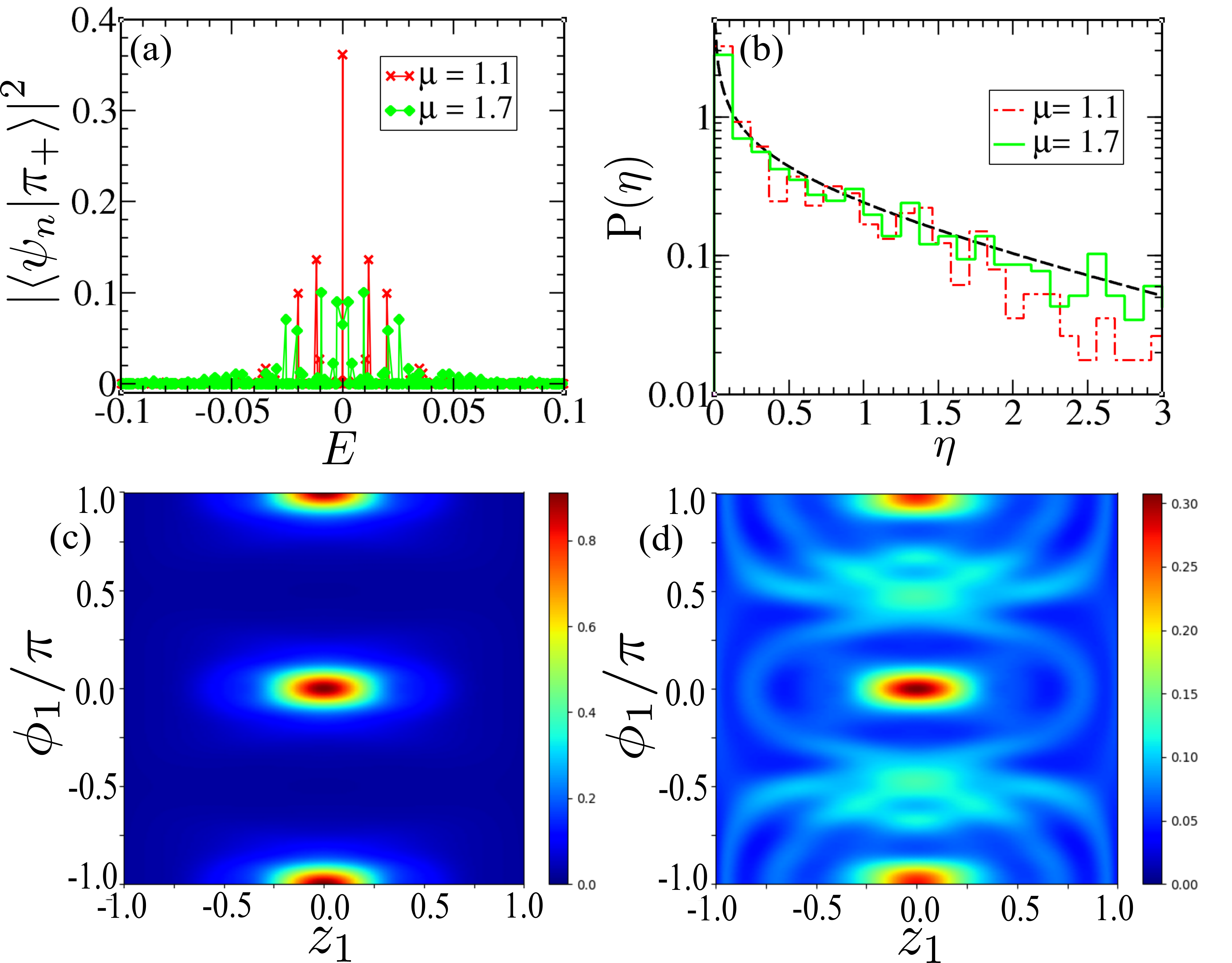}
   \caption{(a) Overlap $|\bra{\psi_n}\ket{\pi_+}|^2$ corresponding to the eigenstates $\ket{\psi_n}$ around the energy density $E=0$. (b) Distribution $\text{P}(\eta)$ corresponding to the eigenstate having maximum overlap near $E\approx0$ for two different values of $\mu$. The black dashed line in (b) denotes the PT distribution. 
Husimi distribution of corresponding eigenstates (shown in (b)) for (c) $\mu=1.1$ and (d) $\mu=1.72$.}
    \label{fig:3}
\end{figure}

To further investigate the degree of chaos at quantum level, we sort the eigenvalues $\mathcal{E}_n$ in ascending order and study the distribution $\text{P}(\delta)$ of the level spacing $\delta_n=\mathcal{E}_{n+1}-\mathcal{E}_n$ corresponding to same sector of parity and exchange symmetries following the usual prescription \cite{Haake,Mehta}. According to BGS conjecture, the level spacing distribution of classically chaotic system follows Wigner-Dyson (WD) statistics \cite{BGS}, whereas Poissonian statistics $\text{P}(\delta) = \exp(-\delta)$ is observed in regular (integrable) regime \cite{Berry_Tabor}. For weak coupling ($\mu\ll\mu_c$), the level spacing follows Poissonian distribution as evident from Fig.\ref{fig:3}(c), whereas in the ergodic regime within a range of intermediate coupling strength, it approaches to WD statistics $\text{P}(\delta) = \frac{\pi}{2}\delta\exp(-\frac{\pi}{4}\delta^2)$ corresponding to Gaussian orthogonal ensemble (GOE) of random matrix (see Fig.\ref{fig:3}(d)). In addition, the average value of ratio of level spacing $\langle r \rangle = \langle \text{min}(\delta_n,\delta_{n+1})/\text{max}(\delta_n,\delta_{n+1}) \rangle$ \cite{note2,Bogomonly_2,David_Huse} confirms such crossover of statistics. Also, similar models with kicked spins exhibit enhanced chaotic behavior with increasing kicking strength \cite{bandyopadhyay,kicked_top_haake}. Apart from spectral statistics, a detailed analysis of eigenvectors provide more information related to ergodicity.

{\it Quantum scars:} Existence of dynamically unstable `$\pi$-mode' (FP-V) within the maximally ergodic regime around the energy density $E=0$ (See Fig.2(b)), motivates us to search for QS above the QPT where the onset of chaos is detected. We investigate the statistical properties of the wavefunction in the said region, which contains more information compared to EE in the present scenario. Our analysis reveals two types of scarred states in the said region.

First type of QS forms due to the unstable `$\pi$-mode', which can be described semiclassically by a symmetric combination of the spin coherent states corresponding to the said FPs,
\begin{equation}
\ket{\pi_+} = \frac{1}{\sqrt{2}}\left(\ket{0,0}\otimes\ket{0,\pi}+\ket{0,\pi}\otimes\ket{0,0}\right)
\label{pi_state}
\end{equation}
where $\ket{z,\phi}$ represents the spin coherent state describing the classical orientation \cite{Radcliffe}. The eigenstates $\ket{\psi_n}$ with scars of `$\pi$-mode' can be identified from the large overlap $|\bra{\psi_n}\ket{\pi_+}|^2$ with the state $\ket{\pi_+}$ representing the `$\pi$-mode' (see Fig.\ref{fig:3}(a)). The Husimi distribution $Q(z,\phi)=\frac{1}{\pi}\bra{z,\phi}\hat{\rho}_\mathcal{S}\ket{z,\phi}$ corresponding to the reduced density matrix $\hat{\rho}_\mathcal{S}$ of such states shows large density at $\phi=0,\pm\pi$ (see Fig.\ref{fig:3}(c),(d)), which is direct evidence of scarring due to `$\pi$-mode'. 
Next, we investigate statistical properties of the wavefunction of scarred states $\ket{\psi_n} = \sum^{\mathcal{N}}_{i=1}\psi^{i}_{n}\ket{i}$ and compute the probability distribution $\text{P}(\eta)$ of the scaled elements $\eta=|\psi^{i}_{n}|^2\mathcal{N}$ to keep the average of $\eta$ to be unity \cite{Haake}. The elements of the eigenvectors of GOE matrices follow Porter-Thomas (PT) distribution $\text{P}(\eta) = (1/\sqrt{2\pi\eta})\exp(-\eta/2)$ \cite{Haake}, which is in accordance with Berry's conjecture describing the Gaussian distribution of amplitudes $\psi^{i}_n$ for higher energy density eigenstates of a classically chaotic  system \cite{Berry}. As seen from Fig.\ref{fig:3}(b), corresponding distribution $\text{P}(\eta)$ for scarred eigenstates deviate from Porter-Thomas distribution indicating the violation of Berry's conjecture which also signifies athermal behaviour \cite{Luitz}. Also comparing Fig.\ref{fig:3}(c) and (d), we observe that the degree of scarring reduces with decreasing overlap $|\bra{\psi_n}\ket{\pi_+}|^2$ and P($\eta$) approaches Porter-Thomas distribution as the system enters into the ergodic regime for increasing $\mu$ (See Fig.3(b)).

\begin{figure} [H]
    \centering
    \includegraphics[height=11.3cm,width=9.05cm]{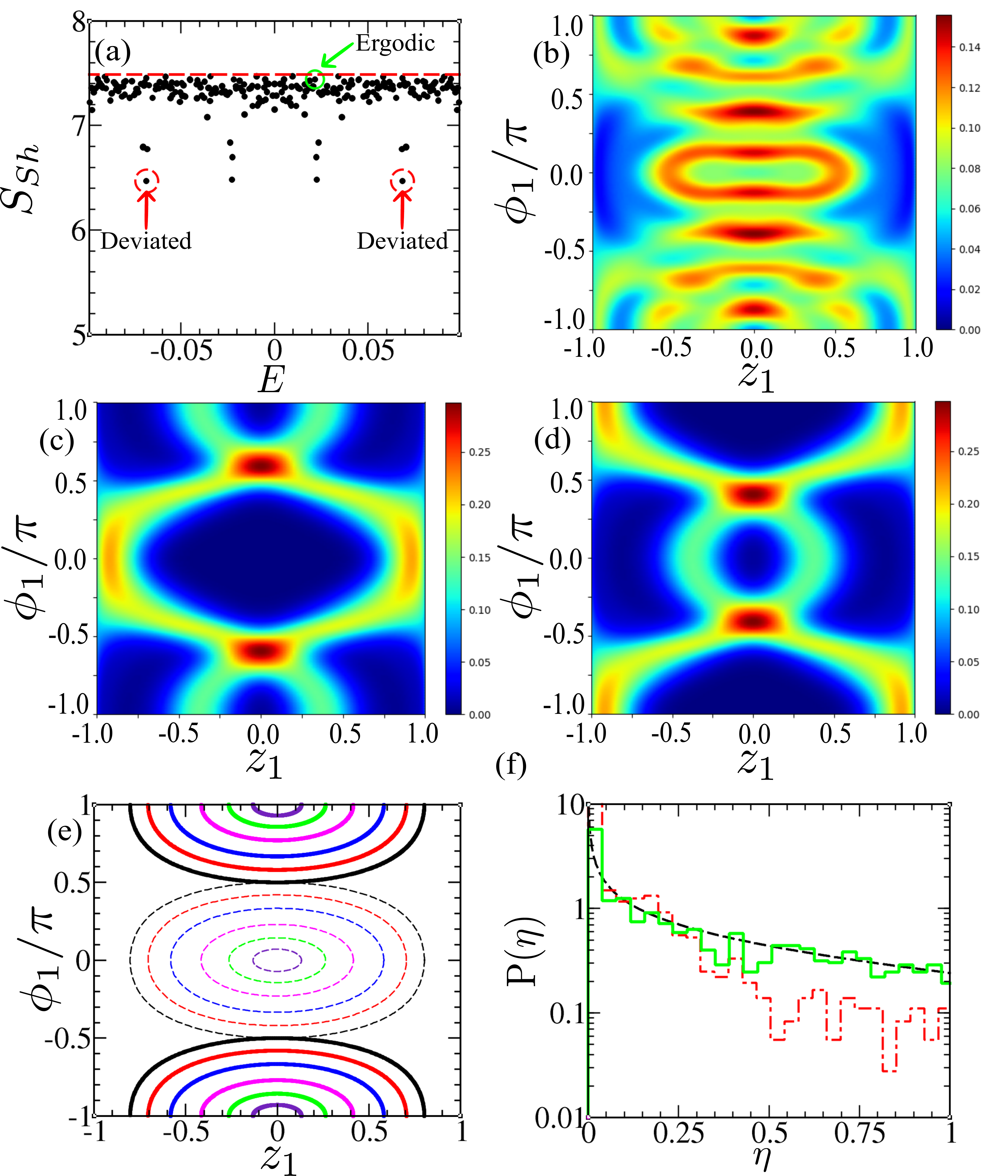}
   \caption{(a) SE of eigenstates around energy density $E=0$. The horizontal red dashed line indicates the GOE limit. Husimi distribution for: (b) a typical ergodic state (shown by green arrow in (a)) and for deviated states: (c) and (d) (indicated by red arrows in (a)). (e) Unstable orbits corresponding to class i (dashed lines) and ii (solid lines). Outermost orbits corresponds to energy density $E=0$. (f) Comparison of distribution $\text{P}(\eta)$ for the deviated states (red dashed-dotted line) and a typical ergodic state (green solid line). The dashed line denotes the PT distribution. Parameter chosen: $\mu=1.85$.}
    \label{fig:4}
\end{figure}

Second type of scarred states are identified from the Shannon entropy (SE), $S_{Sh}=-\sum_{i}|\psi^{i}_{n}|^2\text{log}|\psi^{i}_{n}|^2$ of the eigenstates $\ket{\psi_n}$ around $E=0$ within the ergodic regime near $\mu \approx 2$, where EE approaches the upper limit set by Eq.(\ref{max_EE}) (See Fig.\ref{fig:2}(b)). The SE of most states form a band like structure close to the GOE limit $\text{log}(0.48\mathcal{N})$ \cite{Izrailev1,Izrailev2}; however a few states strongly deviate from it, as shown in Fig.\ref{fig:4}(a). The Husimi distribution of most deviated states reveal high degree of scarring as depicted in Fig.\ref{fig:4}(c),(d). Second type of scars resemble the shape of periodic orbits and are quite different from scar of `$\pi$-mode'. As observed earlier, these scarred states also violate Berry's conjecture and the distribution of elements of the wavefunction deviates from GOE limit (see Fig.\ref{fig:4}(f)). We notice that weakly visible scars can also be present in Husimi distribution of ergodic states in the upper band (see Fig.\ref{fig:4}(b)), however the statistical properties of the wavefunction follow the GOE limit due to much lower degree of scarring compared to the deviated states. 

Next, we search for the origin of second type of QS from the underlying classical dynamics. From Husimi distribution, we observe that the structure of scar remains same for both the spin space resembling the shape of closed orbits with reflection symmetry. Also, the classical Hamiltonian Eq.(\ref{class_Ham}) remains invariant under the transformation $\phi_1 \leftrightarrow \pm\phi_2$ and $z_1 \leftrightarrow \pm z_2$,
which leads us to search for integrable motion within two symmetry classes (i): $\{z_1=-z_2,\phi_1=-\phi_2\}$ or (ii): $\{z_1=z_2,\phi_1=\phi_2\}$. Redefining the collective coordinates as $z_{\pm}=(z_{1} \pm z_{2})/2$ and $\phi_{\pm}=(\phi_{1}\pm\phi_{2})/2$, the condition $\{z_{+}=0,\phi_{+}=0\}$ ($\{z_{-}=0,\phi_{-}=0\}$) corresponding to symmetry class i(ii), remains as steady state and the dynamics of remaining two coordinates is effectively reduced to that of Lipkin Meshkov Glick (LMG) \cite{LMG} model with single spin. For a given energy density $E$, the classical orbits of LMG model can be obtained analytically \cite{shenoy,sm}, as shown in Fig.\ref{fig:4}(e) for different values of $E$ corresponding to two symmetry classes. For $E\approx 0$, the trajectories resemble the shape of QS, moreover the orbits of two symmetry classes touch at $z =0, \phi=\pm \pi/2$, where the Husimi distribution shows large phase space density. The time period of such orbits with $E\approx0$ is given by $T=4K(k)/(1+\mu^2)^{1/4}$ \cite{sm}, where $K(k)$ is complete elliptic integral of the first kind \cite{gradshteyn} with $k^2=\frac{1}{2}(1-1/\sqrt{1+\mu^2})$.

We also study the stability of these orbits using the method of monodromy matrix \cite{monodromy1,monodromy2} and find instability of the orbits for $\mu\geq 1.23$, where Lyapunov exponent increases with $\mu$ \cite{sm}. In this region, we also observe for initial values slightly violating the symmetry condition (i) or (ii), the trajectories deviate from the closed orbit and diffuse in phase space indicating instability. This confirms that the second type of QS is a signature of unstable orbits preserving the symmetry classes. We notice that the degree of scarring reduces with increasing dynamical instability; as a result the deviated scarred states disappear as $\mu$ increases.

At this point, we emphasize that this  model exhibits integrable behavior in two extreme limits of coupling $\mu$, which is manifested by Poissonian level spacing distribution for $\mu\ll 1$, whereas the interaction term in the Hamiltonian (Eqs.(\ref{coupled_top_model}),(\ref{class_Ham})) dominates for $\mu \gg 1$ leading towards integrability \cite{sm} and resulting in a deviation from ergodicity.
   
\begin{figure} 
    \centering
    \includegraphics[height=3.65cm,width=8.5cm]{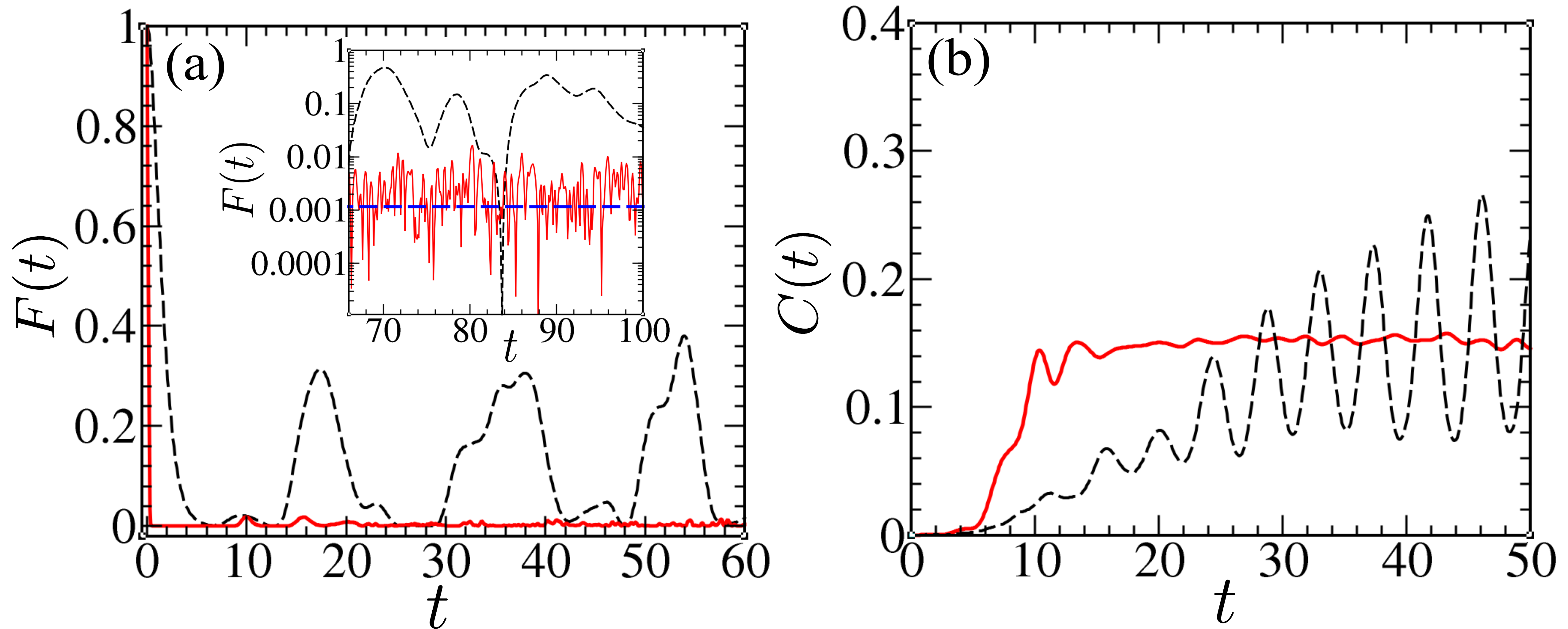}
   \caption{(a) Time evolution of $C(t)$ corresponding to two different initial density matrix $\hat{\rho}_{\text{dev}}$ (black dashed line) and $\hat{\rho}_{\text{mc}}$ (red solid line) for $\mu=1.9$. (b) Comparison of survival probability $F(t)$ corresponding to the initial state `$\pi$-mode' $\ket{\pi_+}$ (black dashed line) and an arbitrary coherent state (red solid line) with energy density $E\approx 0$ for $\mu = 1.3$. Inset shows the long time behaviour of $F(t)$ in semi log scale and horizontal blue dashed line denotes the GOE limit $3/\mathcal{N}$ \cite{Izrailev2,Santos1}. Parameter chosen: $S=25$.}
    \label{fig:5}
\end{figure}

{\it Dynamics of scarred states:} 
To investigate the dynamical signature of two types of QS as mentioned above, we consider two different methods. We study the time evolution of $\ket{\pi_+}$ state (Eq.(\ref{pi_state})) representing first type of QS and compare it with the dynamics of arbitrary coherent state with similar energy density $E\approx0$. The survival probability $F(t) = |\bra{\psi(t)}\ket{\psi(0)}|^2$ computed for $\ket{\pi_+}$ state shows an oscillatory behaviour and significant deviation from the GOE limit $3/\mathcal{N}$ at long time \cite{Izrailev2,Santos1} in contrast to other states (see Fig.\ref{fig:5}(a)), thus indicating non-Markovian dynamics. Even after long time, the Husimi distribution of initial $\ket{\pi_+}$ state retains the QS of `$\pi$-mode' \cite{sinha}.

Next, we analyze the dynamics of the second type of scarred states by studying the ``out-of-time-order correlator" (OTOC), which has recently become an useful tool to diagnose many body quantum chaos \cite{shenker,maldacena,murata_review,Rozenbaum,A_M_Rey1,A_M_Rey2,ising_chain,saturation}. 
Here we investigate the dynamics of unequal time commutator $C(t) = \text{Tr}\hat{\rho} [\hat{W}(t),\hat{V}(0)]^{\dagger}[\hat{W}(t),\hat{V}(0)]$, which is related to OTOC \cite{note3} for unitary $\hat{W}$, $\hat{V}$ and first introduced in the context of superconductivity \cite{Larkin}. In the present case, we consider $\hat{W}=\hat{V}=\hat{S}_{iz}/S$ to study the corresponding $C(t)$. To distinguish the second type of scarred states, we construct two initial density matrices $\hat{\rho}(0)$ namely,
$\hat{\rho}_{\text{mc}}\equiv\sum_{j}\ket{\psi_j}\bra{\psi_j}/\mathcal{N}_{\Delta}$ representing micro canonical ensemble of $\mathcal{N}_{\Delta}$ ergodic states within the small window of energy density $\Delta E \sim 0.1$ around $E=0$ and
$\hat{\rho}_{\text{dev}} = \ket{\psi_{n}}\bra{\psi_{n}}$ corresponding to the deviated state $\ket{\psi_{n}}$ with scar (see Fig.\ref{fig:4}(c)). 
As seen from Fig.\ref{fig:5}(b), for ergodic states, $C(t)$ grows at a faster rate compared to $\hat{\rho}_{\text{dev}}$ and eventually saturates \cite{saturation}, whereas it exhibits oscillations for $\hat{\rho}_{\text{dev}}$ reflecting non-ergodic behavior due to QS. Moreover, the period of such oscillation matches with the time period $T$ of classical orbit reflecting its underlying connection with QS.

{\it Conclusion:} Apart from QPT and dynamical transition, the coupled top model exhibits interesting ergodic behaviour, which we also explored from their connection with underlying classical dynamics. Within an intermediate coupling strength above QPT, the onset of chaos is detected from the dynamics as well from spectral statistics. In the chaotic regime, we identify states with two types of quantum scars arising from unstable steady state and unstable orbits corresponding to symmetry preserved integrable motion. Although, the entanglement entropy of the states at the band center attains the maximum limit, the scarred states lead to the deviation from ergodicity, violating Berry's conjecture. Dynamically, such scarred states can be distinguished from the ergodic states by the saturation value of survival probability as well as oscillations in OTOC, reflecting the periodicity of unstable orbits.

In conclusion, we identified quantum scars as reminiscence of underlying unstable collective dynamics, even in presence of interaction. Such scarred states can be diagnosed from the statistical properties as well from dynamical behavior, exhibiting a clear deviation from ergodicity.\\

{\it Acknowledgements:} We thank Diptiman Sen and Sayak Ray for fruitful discussion.


\begin{thebibliography}{99}
\bibitem{note1}
The exchange operator $\hat{\mathcal{O}}$ is obtained by flipping the indices of basis states $\ket{m_{1z},m_{2z}}$ which reflects the exchange symmetry between the spins.
\bibitem{note2}
The analytical value of $\langle r \rangle$ for Poissonian (GOE) distribution is $\langle r \rangle =2\text{log}2-1 \sim 0.386 \,\,(4-2\sqrt{3}\sim 0.536)$.
\bibitem{note3}
The relation between $C(t)$ and OTOC $\mathcal{F}(t)$ is $C(t) = 2[1-\Re\mathcal{F}(t)]$, where $\mathcal{F}(t)=\text{Tr}\hat{\rho}\hat{W}^{\dagger}(t)\hat{V}^{\dagger}(0)\hat{W}(t)\hat{V}(0)$ for unitary $\hat{W}, \hat{V}$ (See Ref. \cite{shenker,maldacena}) . However, this relation does not hold good for non unitary $\hat{W}=\hat{V}=\hat{S}_{iz}/S$ with $S\gg1$.
\end{thebibliography}
\end{document}


\begin{widetext}

\begin{center}
{\bf SUPPLEMENTARY MATERIAL:\\
 Chaos and quantum scars in coupled top model}
\end{center}

\section{Quantum phase transition and Dynamical transition}
The coupled top model undergoes a quantum phase transition (QPT) as well as a dynamical transition (DT) at the critical coupling $\mu _c = 1$ which is evident from the bifurcation of steady states obtained from semiclassical analysis as given in the main text.  
Quantum mechanically, both QPT and dynamical transition can be identified from the change in the relevant physical quantities associated with the ground state (GS) and highest excited state (ES). Both dynamical transition and QPT are related under the transformation $\hat{S}_{x}\rightarrow -\hat{S}_{x}$ and $\mu \rightarrow -\mu$ of the Hamiltonian changing the ES to GS of anti-ferromagnetic coupled top model which is reflected from the energy density, averages of $\hat{S}_{ix}/S$ and $\hat{S}_{1z}\hat{S}_{2z}/S^2$ as shown in Fig.\ref{fig:1}. Here we point out that the magnetization along the z-axis of individual spins in the broken symmetry sector can not be  obtained directly from the parity conserving eigenstates, however both QPT and dynamical transition are distinctly visible from the sharp change in the quantity $\langle \hat{S}_{1z} \hat{S}_{2z}/S^2\rangle $ with increasing coupling strength $\mu$ as well as ferromagnetic and anti-ferromagnetic spin ordering of respective symmetry broken phase (see Fig.\ref{fig:1}(c),(f)). When spin operators are scaled by its magnitude $S$, the physical quantities converge to the semiclassically obtained analytical result for increasing value of $S$ revealing sharp change at the critical point as shown in Fig.\ref{fig:1}.

\begin{figure}[H]
    \centering
    \includegraphics[height=9cm,width=16.5cm]{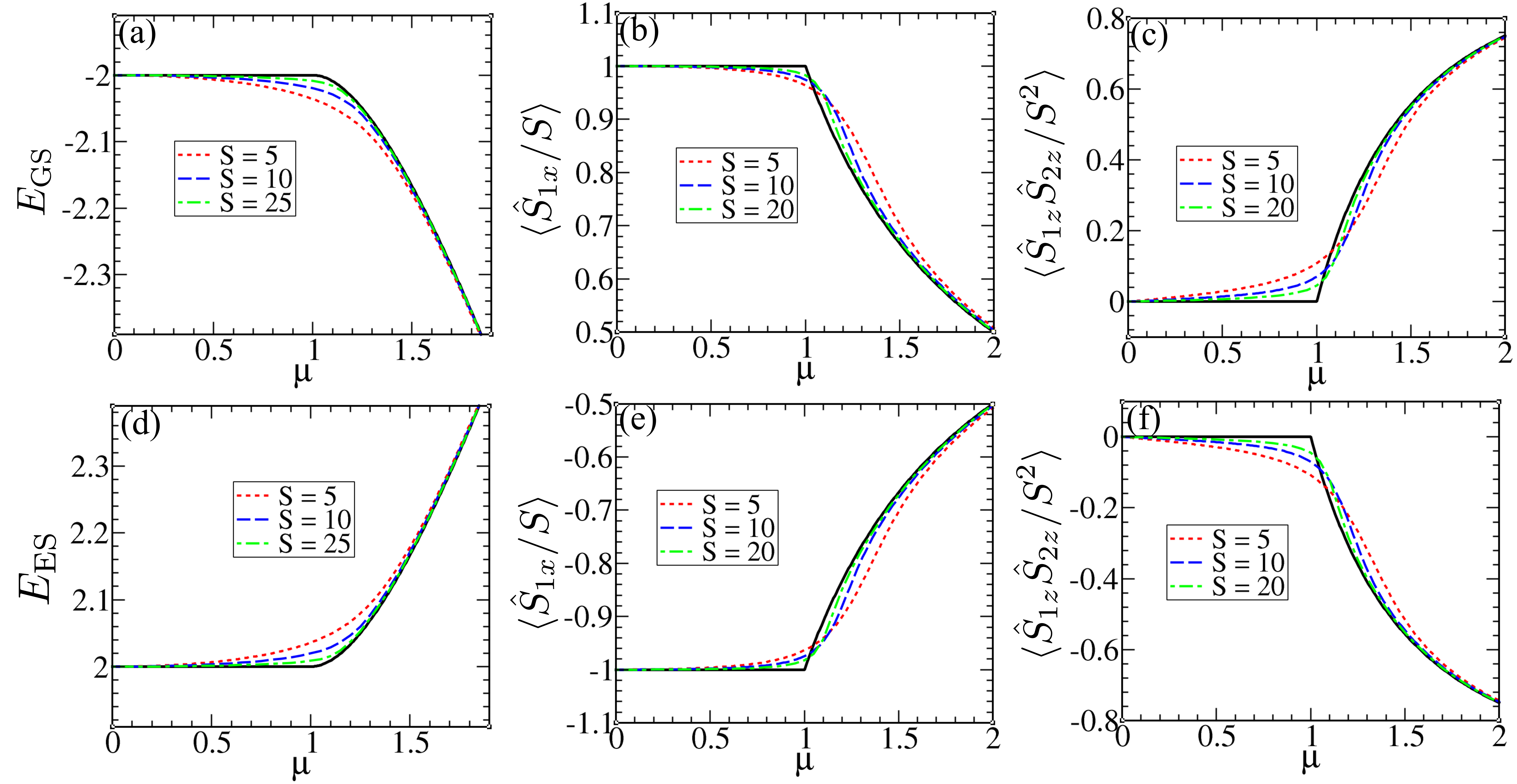}
    \caption{Signature of QPT is captured as a function of coupling strength $\mu$ by (a) energy density $E_{\text{GS}}$ corresponding to ground state, (b) $\langle \hat{S}_{1x}/S \rangle$, c) $\langle \hat{S}_{1z}\hat{S}_{2z}/S^2 \rangle$ whereas that of DT is evident from (d) energy density $E_{\text{ES}}$ corresponding to highest excited state, (e) $\langle \hat{S}_{1x}/S \rangle$  and (f) $\langle \hat{S}_{1z}\hat{S}_{2z}/S^2 \rangle$ for different values of spin $S$. The black solid line in all the figures denote the semiclassically obtained analytical result of the respective quantities.}
    \label{fig:1}
\end{figure}
\begin{itemize}

\item 
{\bf Oscillation frequency:}  
We perform linear stability analysis of the fixed points (FP) obtained from the equations of motion (EOM) (Eq.(3) of the main text) of the collective coordinates $\mathbf{X}=\{z_1,\phi_1,z_2,\phi_2\}$. From the time evolution of small fluctuation $\delta \mathbf{X}(0)$ around the steady state $\mathbf{\bar{X}}$, we obtain the frequency of small amplitude oscillation $\omega$ and the stability of FP is ensured from the condition $\Im(\omega)=0$. The oscillation frequency corresponding to a steady state $\mathbf{\bar{X}}$ can be written as,
\begin{eqnarray}
{\omega}^2&=&\frac{1}{2}\left(\text{A}_1+\text{A}_2-\sqrt{\left(\text{A}_1-\text{A}_2\right)^2+4\mu ^2\frac{\cos\bar{\phi}_1 \cos\bar{\phi}_2}{\sqrt{\text{A}_1\text{A}_2}}}\right)
\end{eqnarray}
where $\text{A}_i=1/(1-{\bar{z}_i}^2)$ where $i = 1,2$. For both QPT and dynamical transition, the excitation frequencies of symmetry unbroken and broken phases are given respectively,
\begin{eqnarray}
\omega &=&
\begin{dcases*} 
  \sqrt{1-\mu} & for  $\mu < 1$ \\ 
  \sqrt{\mu^2-1} & for $\mu \ge 1$
  \end{dcases*} \label{Excitation frequency}
\end{eqnarray} 
which becomes gapless at $\mu=\mu_c$ and vanishes as $\sim|\mu-\mu_c|^{1/2}$ indicating mean field like behaviour of the transitions. Also, the instability of the `$\pi$-mode' (FP-V of the main text) that is the imaginary part of the excitation frequency is given by,
\begin{eqnarray}
\Im(\omega) &=&\sqrt{\frac{(\mu^2+1)^{1/2}-1}{2}}  \label{Excitation frequency-pi}
\end{eqnarray} 
\end{itemize}

\section{Excited state quantum phase transition (ESQPT)}
\noindent The coupled top model exhibits excited state quantum phase transition (ESQPT) above the critical coupling $\mu_c = 1$  corresponding to the unstable symmetry unbroken FPs I and II with energy density $E$ = -2 and 2 respectively. The ESQPT is associated with the singularity of semiclassical density of states (DOS) at critical energy densities \cite{Cejnar_1,Caprio,Cejnar_2,dicke_ESQPT,LF_Santos1}.  
Using the semiclassical Hamiltonian $\mathcal{H}_{cl}$ ( Eq.(2) in main text) the DOS can be written as, 
\begin{eqnarray}
\rho(E) &=& \frac{\mathcal{C}}{(2\pi)^2}\int_{-1}^{1}\int_{-1}^{1}\int_{0}^{2\pi}\int_{0}^{2\pi}\delta(E-\mathcal{H}_{cl})\,d\phi_1\,d\phi_2\,dz_1\,dz_2
\label{DOS}
\end{eqnarray}
where we use the normalization condition $\int \rho(E)\,dE = 1$, which yields $\mathcal{C}=1/4$. We also calculate DOS from quantum mechanical spectrum which is in good agreement with semiclassical result as compared in Fig.\ref{fig:2}(a). For $\mu>\mu_c$, the derivative of the semiclassical DOS shown in the inset of Fig.\ref{fig:2}(a) reveals the singularity at critical energy densities $E = \pm 2$. Physically above $\mu_c$, the symmetry unbroken FPs I and II at critical energy densities $E=\pm 2$ separates symmetry unbroken states within the range $-2<E<2$ from symmetry broken states $E_{GS} < E < -2$ and $ 2 <E < E_{ES}$ corresponding to QPT and dynamical transition respectively.  As shown in Fig.\ref{fig:2}(b), the pair gap defined as $\Delta_{n} =\mathcal{E}_{2n}-\mathcal{E}_{2n-1}$ ($n=1,2,...$) vanishes exponentially with $S$ for $E<-2$ and $E>2$, revealing symmetry broken nature of quasi degenerate consecutive even odd parity states. 
The average values of the observables like $\langle \hat{S}_{1z}/S\rangle$ and $\langle \hat{S}_{1x}/S\rangle$ clearly distinguish symmetry unbroken states from symmetry broken sector with a significant change at $E = \pm2$, indicating ESQPT (see Fig.\ref{fig:2}(c),(d)).

\begin{figure}[H]
    \centering
    \includegraphics[height=8.2cm,width=10.3cm]{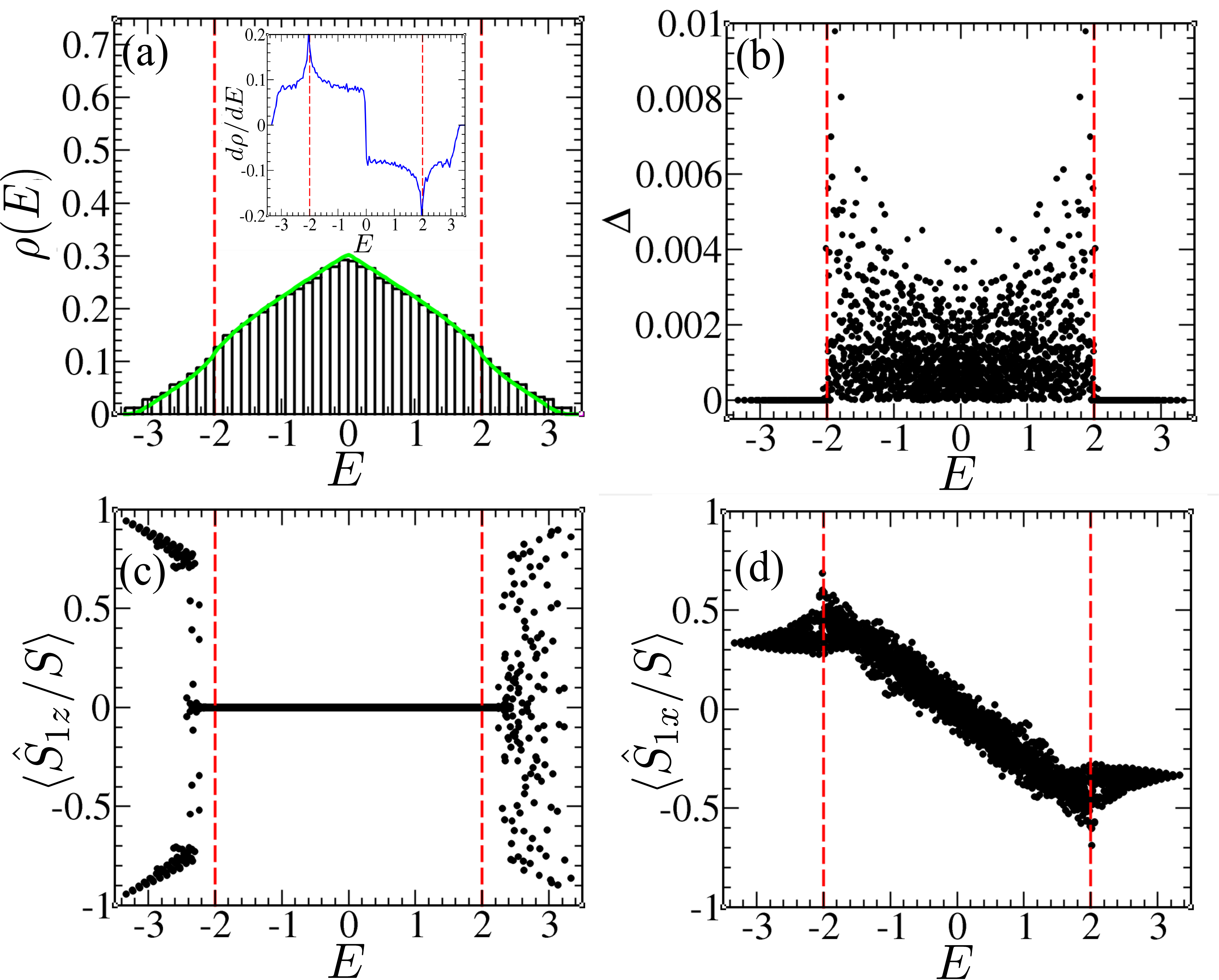}
    \caption{Manifestation of ESQPT using (a) $\rho(E)$ and its derivative $d\rho /dE$ given in the inset, (b) pair gap $\Delta$, (c) $\langle\hat{S}_{1z}/S\rangle$ and (d) $\langle\hat{S}_{1x}/S\rangle$ as a function of energy density $E$ at $\mu=3$. The red dashed lines marked at $E=\pm2$ indicate the critical energy densities corresponding to ESQPT. }
    \label{fig:2}
\end{figure}
\section{Ergodicity and entanglement entropy at different energy densities}
\begin{figure}[H]
    \centering
   \includegraphics[height=10.8cm,width=16.5cm]{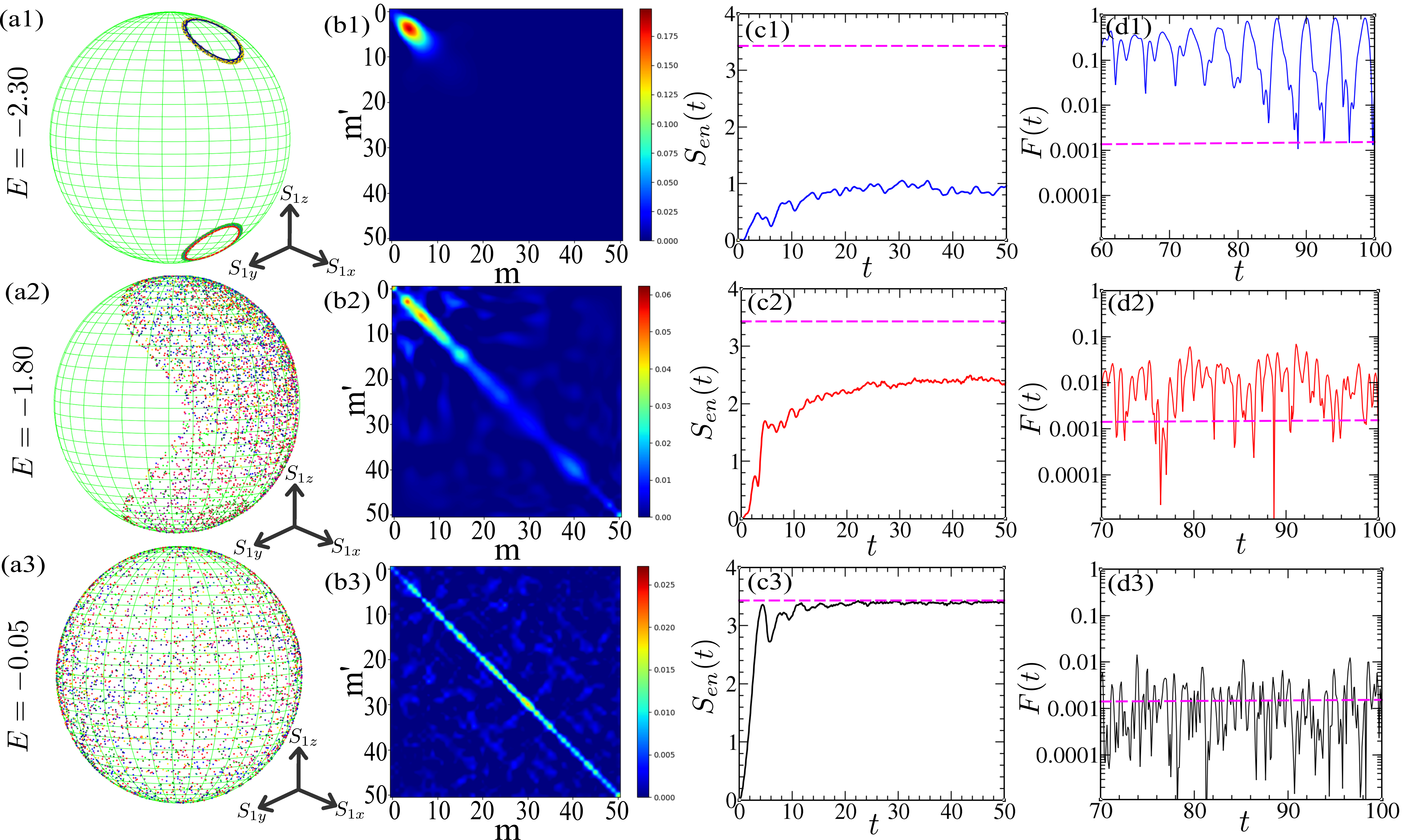}
    \caption{Connection between ergodicity and entanglement entropy (EE) with increasing energy density $E$ (as shown in different rows). Column(a): The classical phase space trajectories on Bloch sphere with fixed classical energy $E$. Column(b): color scale plot of the elements of final reduced density matrix $(\hat{\rho}_{\mathcal{S}})_{mm'}$ (in the basis of $\hat{S}_z$), after time evolution of an initial coherent state with energy density $E$. Column(c) and (d) represent the time evolution of EE and survival probability respectively for the same initial coherent states. Horizontal dashed lines in column(c) and (d) denote maximum limit of EE (Eq.(4) of the main text) and  GOE limit of survival probability $3/\mathcal{N}$ respectively. Parameters chosen: $\mu=1.85$ and $S=25$.}
    \label{fig:3}
\end{figure}
 In the intermediate coupling strength ($\mu \approx 2$), where the onset of chaos is detected from the change in spectral statistics, it is observed that the entanglement entropy (EE) of eigenstates increases with ascending energy density up to the band center with $E\approx 0$ (as shown in Fig.2(b) of the main text) and the peak value of EE reaches the maximum limit $S_{\text{max}}$ (Eq.(4) of the main text). Such behavior of EE indicates enhanced degree of ergodicity with increasing energy density, which we investigated from both classical and quantum dynamics as summarized in Fig.\ref{fig:3}. Below ESQPT ($E<-2$), when the classical energy is close to the ground state, the trajectories are confined within a small region around the symmetry broken fixed points (FP-III), clearly exhibiting a localized structure in phase space as shown in Fig.\ref{fig:3}(a1). Above the ESQPT with increasing energy density $E$, the diffusive behavior of phase space trajectories are observed which occupy the Bloch sphere partially (see Fig.\ref{fig:3}(a2)) and finally, they cover the full Bloch sphere for $E\approx0$ (see Fig.\ref{fig:3}(a3)), which also corresponds to energy density of the states at the band center where EE saturates to its maximum limit $S_{\text{max}}$ (Eq.(4) of the main text). For quantum dynamics, we evolve a coherent state with corresponding energy density $E$, and study the properties of reduced density matrix, EE and survival probability at long times. With  increasing energy density $E$, the final reduced density matrix $(\hat{\rho}_{\mathcal{S}})_{mm'}$ approaches to diagonal form with increasing EE as evident from Fig.\ref{fig:3}(b) and 3(c). As seen in Fig.\ref{fig:3}(b3) and 3(c3), for energy density $E\approx 0$, the reduced density matrix takes the diagonal form and corresponding EE reaches the maximum limit $S_{\text{max}}$ (Eq.(4) of the main text) indicating ergodic behavior. Also, the long time behavior of the survival probability $F(t)$ approaches to the GOE limit $3/\mathcal{N}$ with increasing energy density (see Fig.\ref{fig:3}(d1)-(d3)). Note that, the ergodic behavior at energy density $E\approx 0$ can be observed for an arbitrary initial coherent state except the special state $\ket{\pi_+}$ (Eq.(5) of the main text), which leads to the deviation from ergodicity due to the scarring phenomena as discussed in the main text. Such connection between ergodicity in the sense of phase space mixing and EE are also investigated in different contexts \cite{lewenstein,shohini_ghose,EE_expt}. 
\section{Classical periodic orbits}
Due to the symmetry of the Hamiltonian $\mathcal{H}_{cl}$  ($z_{1} \leftrightarrow \pm z_{2}, \phi_{1} \leftrightarrow \pm \phi_{2}$), we explore the integrable motion within two symmetry classes: I $\{z_{1}=-z_{2},\phi_{1}=-\phi_{2}\}$ and II $\{z_{1}=z_{2},\phi_{1}=\phi_{2}\}$. These constraints remain as steady states $\{\phi_+=0;z_+=0\}$ and $\{\phi_-=0;z_-=0\}$ corresponding to symmetry class I and II respectively, where we define new coordinates as $z_\pm = (z_1 \pm z_2)/2$ and $\phi_\pm = (\phi_1 \pm \phi_2)/2$. Using the conditions mentioned above, the remaining degrees will satisfy the EOM of an effective anti-ferromagnetic (for I) or ferromagnetic (for II) LMG model,
\begin{eqnarray}
&&\dot{z}_-=-\sqrt{1-{z_-}^2}\sin\phi_-;\quad
\dot{\phi}_-=\frac{z_-}{\sqrt{1-{z_-}^2}}\cos\phi_-+\mu z_-\quad (\text{For\,class\,I})
\label{equation_z-phi-} \\
&&\dot{z}_+=-\sqrt{1-{z_+}^2}\sin\phi_+;\quad
\dot{\phi}_+=\frac{z_+}{\sqrt{1-{z_+}^2}}\cos\phi_+-\mu z_+\quad (\text{For\,class\,II})  \label{equation_z+phi+}
\end{eqnarray}
The solution of above equations \cite{Shenoy} is given in terms of elliptic function as,
\begin{equation}
z_\pm(t)=C\,\text{cn}\left(\frac{C\mu}{2k}(t+t_0),k\right),~~\cos(\phi_\pm(t))=-\frac{E+\xi \mu \,{z_\pm(t)}^2}{2\sqrt{1-{z_\pm(t)}^2}} \label{jacobi-elliptic-functions}
\end{equation}
where cn is the Jacobi elliptic function with elliptic modulus $k$ and the constants are defined in the following way,
\begin{eqnarray}
C^2\,=\,\frac{2}{\mu ^2}\left[-\xi\frac{\mu E}{2}-1+\Omega\right],\quad k^2&=&\frac{1}{2}\left[1-\frac{\xi E\mu/2 +1}{\Omega}\right],\quad t_0 =\frac{F\left(\cos^{-1} (z_\pm(0)/C),k\right)}{\Omega^{1/2}},\quad \Omega =\sqrt{\mu^2+1+\xi E\mu} \nonumber
\label{elliptic-modulus}
\end{eqnarray}
\noindent The parameter $\xi \mu$ where $\xi=-1 (+1)$ for class I (II) acts as the coupling constant for the effective anti-ferromagnetic (ferromagnetic) LMG as seen from Eq.(\ref{equation_z-phi-}) (Eq.(\ref{equation_z+phi+})). Within the energy density range $-2<E<2$, the elliptic modulus is always $k<1$. Interestingly, the constants $C,k,\Omega$ are same for both the periodic orbits formed close to the unstable FP-I (II) belonging to class I (II) corresponding to energy density $E<0$ ($E>0$), which implies that the shape of the orbits belonging to two different symmetry classes having equal and opposite energy densities are same, but with a $\pi$ shift along $\phi$. The time period $T$ for oscillations of $z_\pm(t)$ is given by,
\begin{eqnarray}
T &=& \frac{8kK(k)}{C\mu} \label{time-period}
\end{eqnarray}
where $K(k)\,=\,F(\pi/2,k)$ is the complete elliptic integral of first kind and $F(\phi,k)\,=\,\int_{0}^{\phi}dx (1-k^2\sin^2 x)^{-1/2}$. To analyze the stability of these orbits, we calculate the Lyapunov exponent (LE). The LE is a measure which characterizes the growth of a small perturbation given to the solution of the system. If the small perturbation $\delta\mathbf{X}(t=0)$ to the collective coordinates $\mathbf{X}=\{z_1,\phi_1,z_2,\phi_2\}$ grows exponentially in time, then one can write $||\delta\mathbf{X}(t)||=e^{\lambda t} \,||\delta\mathbf{X}(0)||$ and characterize the LE ($\lambda$) as,
\begin{eqnarray}
\lambda&=&\lim_{t \to \infty}\frac{1}{t}\ln \left(\frac{||\delta\mathbf{X}(t)||}{||\delta\mathbf{X}(0)||}\right)
\label{definition_lyapunov_exponent}
\end{eqnarray}
\noindent When the limit in Eq.(\ref{definition_lyapunov_exponent}) exists and is positive, the trajectory is extremely sensitive to the initial condition and thus becomes chaotic in nature. 
To obtain LE, we calculate Monodromy matrix (MM) \cite{monodromy1,monodromy2} $\mathbf{M}$ such that $\delta\mathbf{X}(t)=\mathbf{M}(t)\,\delta\mathbf{X}(0)$, and the differential equation governing the evolution of MM is given by,
\begin{eqnarray}
\frac{d\mathbf{M}(t)}{dt}&=&\mathbf{J}(\mathbf{X}(t))\,\mathbf{M}(t)
\label{evolution_of_monodromy_matrix}
\end{eqnarray}
\noindent Where $\mathbf{J}(\mathbf{X}(t))$ is the Jacobian matrix.
In order to obtain the LE for autonomous system, it is necessary
to solve the dynamical equations and the elements of MM $\mathbf{M}(t)$ simultaneously where the initial MM is identity matrix. By evolving the MM upto one time period $T$ given by the Eq.(\ref{time-period}), we obtain the largest eigenvalue of the MM. Then the largest lyapunov exponent (LLE) is given by,
\begin{eqnarray}
\lambda_l=\ln(m_l)/T
\label{LLE}
\end{eqnarray}
where $m_l$ is the largest eigenvalue of MM at $t=T$.

The LLE ($\lambda_l$) of the periodic orbits corresponding to energy density $E=0$ become positive for $\mu>\mu_u = 1.23$ (see Fig.\ref{fig:4}(a)), which shows that the orbits in the ergodic regime around $\mu\approx 2 $ (as discussed in main text) are indeed unstable.
As seen from Fig.\ref{fig:4}(a), the stability region decreases for periodic orbits with increasing magnitude of energy density $|E|$ (from $E=0$) as LLE becomes positive for $\mu<\mu_u$, and finally it terminates at $\mu_c=1$ for $E=\pm 2$ corresponding to unstable symmetry unbroken FPs I and II.
\begin{figure}[H]
    \centering
    \includegraphics[height=8.5cm,width=10cm]{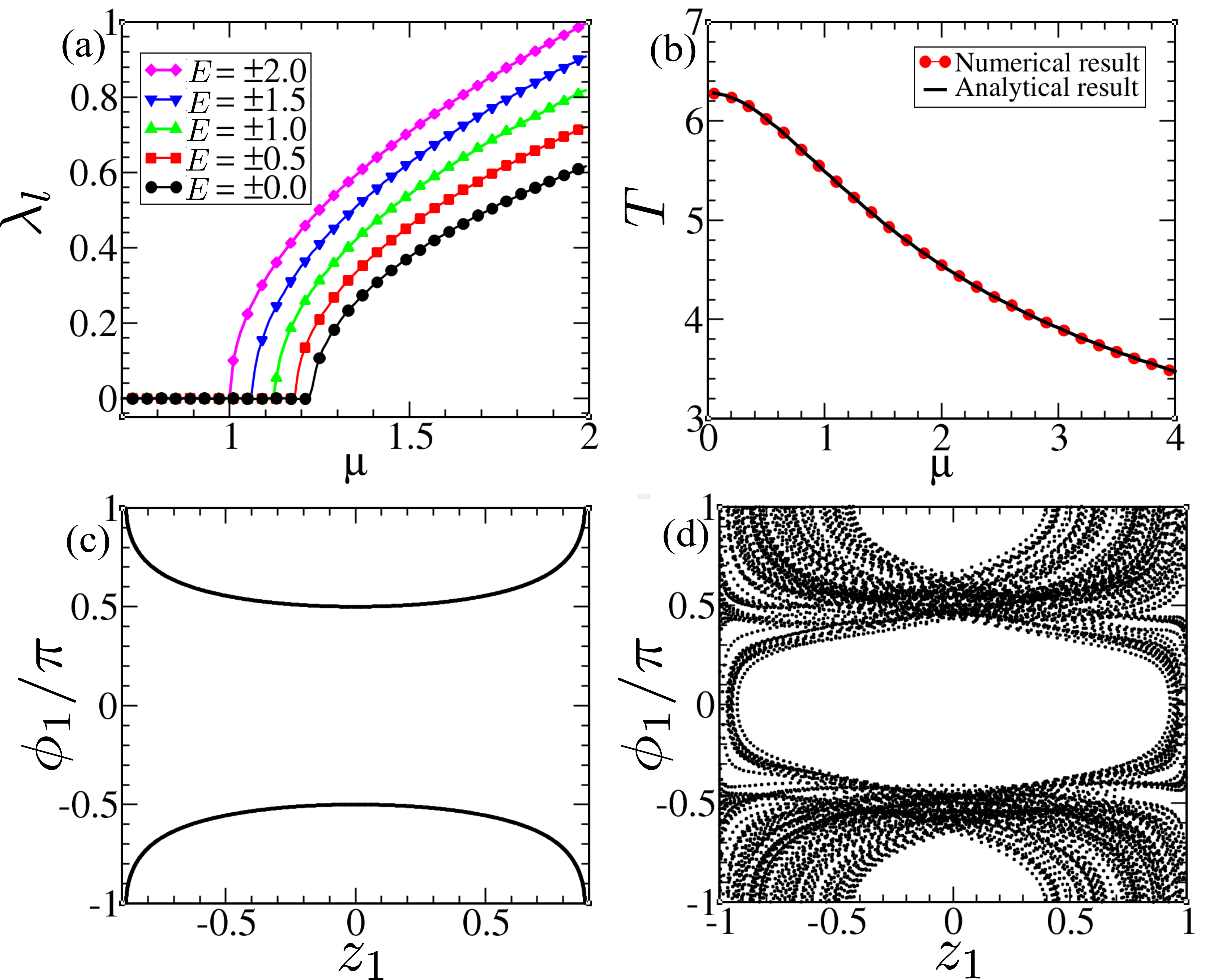}
    \caption{Instability of periodic orbits: (a) LLE ($\lambda_l$) as a function of $\mu$ for the periodic orbits with different energies, (b) time period of periodic orbits corresponding to energy density $E=0$ as a function of $\mu$. Evolution of a periodic orbit  from an initial condition $z_-(0)=\delta z$ corresponding to symmetry class II ($z_-=0$) with energy density $E\approx 0$, where $\delta z\ll 1$ is a small perturbation: (c) in the stable region with $\mu=1.2$ and (d) unstable region with $\mu=1.4$.}
    \label{fig:4}
\end{figure}

Moreover, the LLE for $E =\pm 2$ coincides with the imaginary part of the excitation frequency $\Im (\omega)=\sqrt{\mu-1}$ (Eq.(\ref{Excitation frequency})) corresponding to FP I and II. 
This clearly indicates that the periodic orbits corresponding to class I(II) with energy  density $E<0(E>0)$ formed around the unstable FP-I(FP-II), as shown in Fig.4(e) in the main text.

To confirm the stability of orbits with $E\approx 0$ in dynamics, we study time evolution of an initial point on a periodic orbit under small perturbation deviating from the symmetry class of corresponding orbit. In the stable region $\mu<\mu_{u}$, the periodic orbit is formed even in presence of symmetry breaking perturbation as seen from Fig.\ref{fig:4}(c).
On the contrary, for $\mu > \mu_u$, the trajectory of perturbed initial point diffuses in phase space without forming closed orbit (see Fig.\ref{fig:4}(d)) indicating instability. Although such perturbed trajectory does not form periodic orbit of particular symmetry class in the unstable region, it evolves around the orbits of both the symmetry classes diffusively, as seen from Fig.\ref{fig:4}(d). Interestingly, such phase space structure of unstable orbit resembles the shape of QS observed in Husimi distribution shown in Fig.4(c) of main text. 
  
\section{Integrability at large coupling strength}
It is interesting to note that the model becomes integrable in two extreme limit of coupling strength $\mu$. For $\mu \rightarrow 0$, the Hamiltonian describes two non interacting spins precessing around the x-axis. At the quantum level, this is manifested by Poissonian level spacing distribution for $\mu < 1$. 
In the opposite limit $\mu \gg 1$, we scale the Hamiltonian (Eq.(1) of the main text) by $\mu$ which can be rewritten as,
\begin{eqnarray}
\hat{\mathcal{H}} = -\epsilon(\hat{S}_{1x}+\hat{S}_{2x})-\frac{1}{S}\hat{S}_{1z}\hat{S}_{2z} \label{integrable_limit}
\end{eqnarray} 
where $\epsilon\equiv1/\mu$ and $\epsilon\rightarrow 0$ ($\mu\rightarrow \infty$). 
Semiclassically, the above Hamiltonian can be described by the collective coordinates as,
\begin{eqnarray}
\mathcal{H}_{cl} = -\epsilon\left(\sqrt{1-z_1^2}\cos{\phi_1}+\sqrt{1-z_2^2}\cos{\phi_2}\right)-z_1z_2.
\label{semiclassical_ham_integrable}
\end{eqnarray}
For $\epsilon = 0$, the classical Hamiltonian $\mathcal{H}_{cl}$ is integrable and independent of angle variable.
The EOM yields the solution $\phi_{i} = z_{i}t + C$ with constant $z_{i}$, which represents precession of spins around the $z$-axis as depicted in Fig.\ref{fig:5}.

\begin{figure}[H]
    \centering
    \includegraphics[height=3.7cm,width=8.5cm]{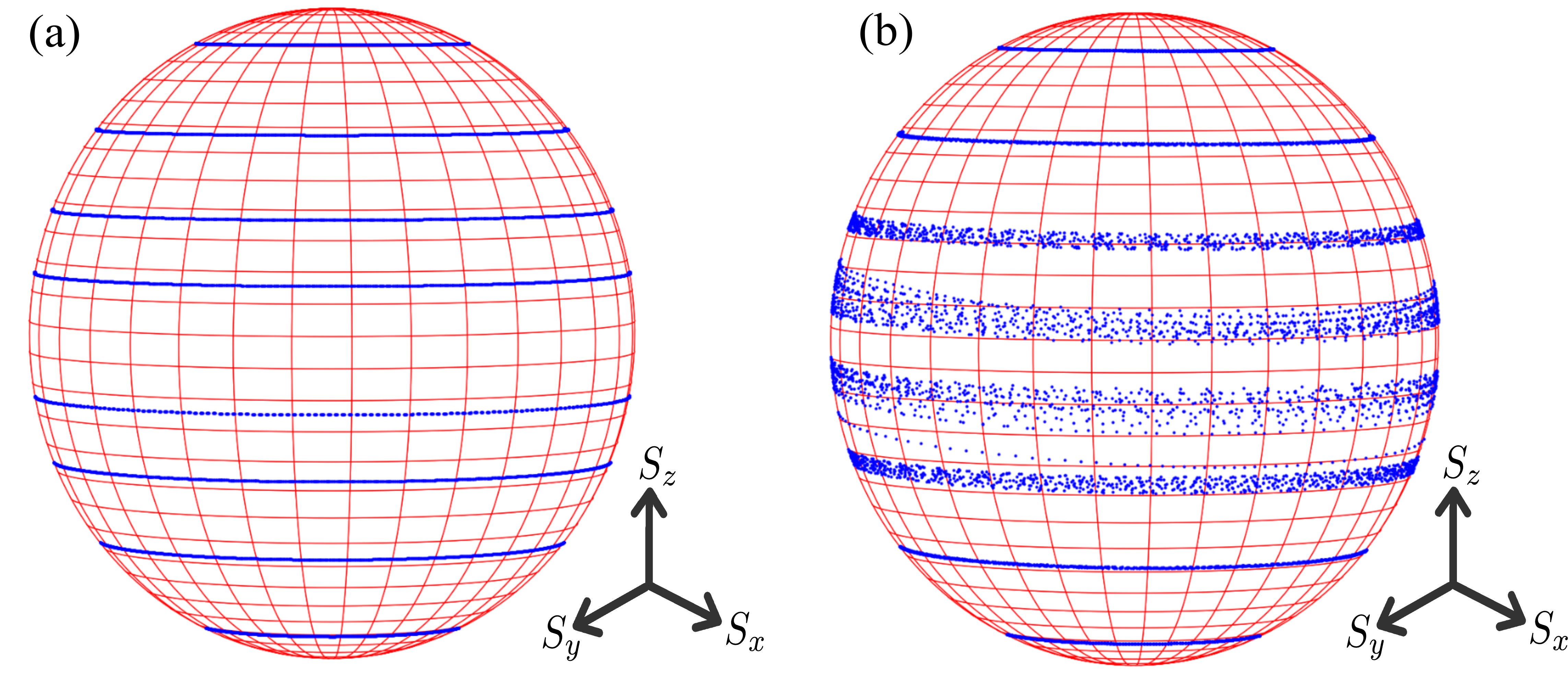}
    \caption{Bloch sphere trajectories shown in a particular spin sector at different energies (a) for $\epsilon=0$ and (b) for $\epsilon=0.01$. Diffusive behaviour in the trajectories is observed for $\epsilon=0.01$ near $E=0$ whereas a regular behaviour is observed away from the equator.}
    \label{fig:5}
\end{figure}

The above results obtained from semiclassical analysis at large $\mu$ are also supported by quantum analysis as well. To see this, we sort the energy levels obtained from exact diagonalization of the above Hamiltonian (Eq.(\ref{integrable_limit})) in ascending order belonging to a particular parity as well as exchange symmetry sector (mentioned in main text). Next, we compute the level spacing $\delta_n=\mathcal{E}_n-\mathcal{E}_{n-1}$, which is shown with increasing value of the index $n$ in Fig.\ref{fig:6}.
For the integrable limit with $\epsilon=0$, the level spacings are distributed in a regular fashion compared to the that for $\epsilon\neq0$. However, for $\epsilon \ll 1$, the level spacings corresponding to very low as well as high energy states show almost regular structure, whereas those at the middle of the spectrum ($E\approx 0$) exhibits significant deviation which is consistent with the classical picture (see Fig.\ref{fig:4}). Also such deviation increases as the Hamiltonian differs from the integrable limit with increasing value of $\epsilon$. A similar analysis has also been done in the context of Dicke model \cite{Dicke}.

\begin{figure}[H]
    \centering
    \includegraphics[height=5.8cm,width=7.4cm]{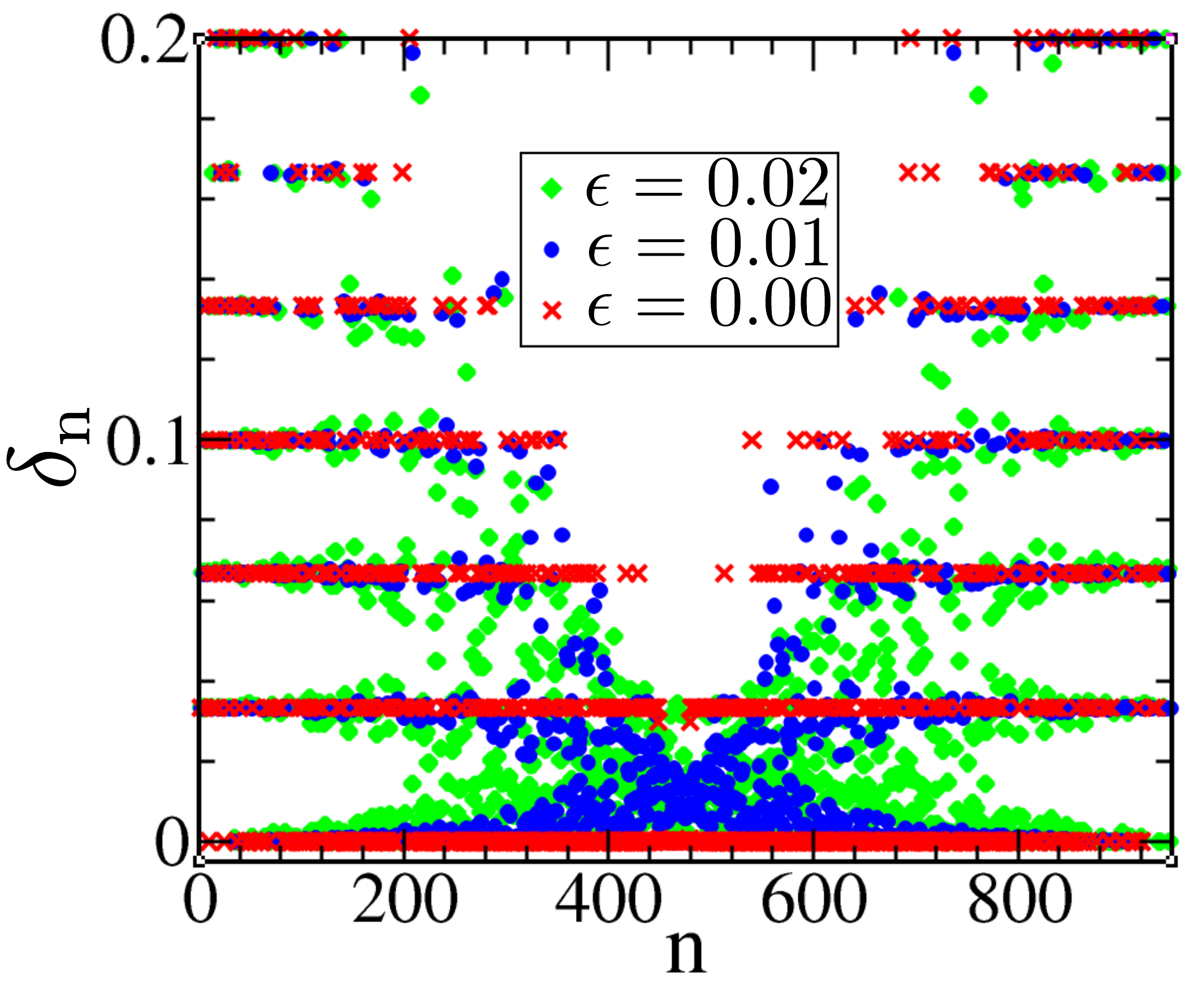}
    \caption{(a) Plot of level spacing $\delta_n$ with index $n$ for different values of $\epsilon$ showing a regular behavior at $\epsilon=0$ and almost regular behavior for very low as well as high energy eigenstates. Parameter chosen: $S=30$. The symmetry sector is considered even for both the symmetry operators.}
    \label{fig:6}
\end{figure}

\end{widetext}